\documentclass[prstab,twocolumn,superscriptaddress]{revtex4}
\usepackage{graphicx}
\usepackage{amssymb}
\usepackage{amsmath}
\usepackage{underscore}
\usepackage{natbib}

\usepackage[caption=false]{subfig}

\usepackage{eurosym}

\makeatletter
\newcommand\MEUR[1]{\if@EURleft\text{\euro}\,\fi#1\if@EURleft\else\,\text{\euro}\fi}
\makeatother

\begin{document}


\title{Cost Optimisation of an Instrument Suite at an Accelerator-Driven Spallation Source}



\author{P.M. Bentley}
\affiliation{European Spallation Source ESS AB, P.O. Box 176, 221 00 Lund, Sweden}
\affiliation{Department of Physics and Astronomy, Uppsala University, Box 256, 751 05 Uppsala, Sweden}
\email[]{phil.m.bentley@gmail.com}

\author{C.S. Zendler}
\affiliation{European Spallation Source ESS AB, P.O. Box 176, 221 00 Lund, Sweden}

\author{O. Kirstein}
\affiliation{European Spallation Source ESS AB, P.O. Box 176, 221 00 Lund, Sweden}
\affiliation{Department of Physics and Astronomy, Uppsala University, Box 256, 751 05 Uppsala, Sweden}
\affiliation{School of Mechanical Engineering, University of Newcastle, Australia}



\date{\today}

\begin{abstract}
  This artcile presents an optimisation of performance and cost of
  neutron scattering instrumentation at the European Spallation
  Source.  This is done by trading detailed cost functions against
  beam transmission functions in a multi-dimensional, yet simple,
  parameter space.  On the one hand, the neutron guide cost increases
  as a power of the desired beam divergence, and inversely with the
  minimum wavelength, due to the supermirror coating needed.  On the
  other hand, the more neutrons are transported to the instrument the
  greater are the shielding costs to deal with the gamma rays that
  result from the eventual absorption of the neutrons.  There are
  additional factors in that many of the parameters defining the
  neutron guide geometry are continuous variables, and the
  straightness of the guide increases the transmission of high energy
  spallation products, which affect the specifications of particularly
  heavy hardware, such as heavy shutters and additional shielding,
  beam stops etc.  Over the suite of 16 instruments, a maximum
  potential cost saving is identified in the vicinity of 20 M\euro \,,
  with negligible penalties on instrument performance and bandwidth,
  by trading minimum neutron wavelength $\lambda$ against total
  optical costs and shielding costs.  Finally, these cost calculations
  are shown to be consistent with those at real world, existing
  spallation source facilities.
\end{abstract}


\maketitle {}

\section{Introduction}

Modern neutron scattering instruments at spallation sources are
expensive, costing in the region of 12-20 M\euro.  Typically, an
instrument has a working lifetime of approximately 10-20 years.  At
the time of writing, the budget for the instrument suite of 16
instruments at the ESS, including ramping up all of the suport
functions, is around 211 M\euro, from a total facility budget of 1\,843
M\euro.

It is imperative that facilities of this scale strike the appropriate
balance of cost and performance, to maximise the return on investment
and create as large a benefit as possible from the available
resources.

It will be shown later in this report that a significant fraction of
the cost of each instrument, around 30-50\%, is spent on shielding and
neutron optical components \cite{ESS-CostCatWorkshop}.  Indeed,
shielding is the main cost driver, costing around 30\% of the total
budget.  Shielding is directly linked with instrument background noise
--- and hence instrument user quality --- along with safety and
regulatory factors.  As such, one might expect that this area of
expertise would receive more attention in the neutron scattering
community than it does.

In the past, shielding and optics tended to be optimised somewhat
independently.  We show in this article that successful cost
optimisation and instrument background reduction is considerably
easier if the whole system is designed and optimised together as a
single task for an expert team.  Indeed, at the European Spallation
Source (ESS), Sweden, and the Paul Scherrer Institut (PSI),
Switzerland, a core risk and cost minimisation strategy is to design
the optics and shielding systems together by co-located personnel
using a common set of tools and with thorough cross-checking in place.

In this article, an initial stage of this work is summarised.
Preliminary specifications and requirements are identified for the
instrument suite beamline systems, and the cost of those systems is
minimised to a critical performance metric value to be defined
shortly.  Systems exceeding this metric value could be defined as
being over-engineered, and systems below this value as
under-performing.  This allows us to define an optimum cost for the
instrument suite.

Finally, it will be shown that by strategically selecting a subset of
``primary'' instruments, which are fully optimised, and reducing the
specifications of the remainder into the under-engineered category, a
further set of cost savings are possible with a viable suite, with a
small decrease in performance.

\section{Input Data and Cost Estimates}
There are two beam scenarios to consider in this initial study.  The
first evaluates the shielding costs for a direct beam with line of sight back
to the source, and the latter is with part of the beam line curved out of line of sight.

\subsection{Line of Sight Option}
This shielding is thick, since it must protect against scattered high
energy (MeV - GeV) neutrons and gamma rays.  We use a rough cost
estimate for an enhanced concrete instrument cave for straight
beamlines, based on the PSI instrument ``BOA'', increased in thickness
to 2 metres, and accomodating $5\times 5\times 5$ m$^3$ of instrument
space.  These specifications were for a generic instrument, based
partly on calculations of minimum thickness for a MW spallation source
with a GeV proton beam, and partly on measurements at PSI and SNS on
fast neutron transmission and prompt pulse backgrounds, which usually adversely affects instrument performance. 

A completely straight beamline also requires a heavy chopper ---
commonly known as ``$T0$'', ``$t$-zero'', or ``Prompt Pulse
Suppression'' (PPS) choppers: these are heavy choppers that attenuate
the high-energy particles during proton illumination of the spallation
target.  The cost of these T0 choppers was estimated roughly and
somewhat optimistically by the ESS Chopper group, assuming some
economy of scale by manufacturing multiple units and minimising
development costs.

The instrument also requires a heavy shutter, capable of blocking the
high energy beam and allowing access to the sample area.  A study
performed by the ESS engineering department for an SNS-style
cylindrical drum shutter was completed in 2013, which provided the
cost estimates for this item.  Note that a guillotine-style shutter
outside the monolith is not recommended due to the vertical streaming
paths this would generate next to the short instruments.

\subsection{Out of Line of Sight}
This option has a similar cost of guide shielding as that of an
instrument at a reactor source.  A thickness of 60 cm of regular
concrete was found via preliminary Monte-Carlo calculations to
attenuate the radiation sufficiently to safe levels (1.5 $\mu$Sv/h
total simulated dose rate \footnote{The requirement is actually
  3$\mu$Sv/h. A factor of 2 between simulated dose rate and actual
  requirement is used to compensate for discrepancies between computer
  models and engineering designs.}). 60 cm of concrete is slightly
thicker than other lower power facilities but not excessive.  For
example, on TS2 at ISIS, 35 cm of concrete shielding is sufficient,
although in that case a heavy shutter can be closed if there is a
safety hazard on a beamline.

\subsection{Concept Neutral Costs}

For the instrument cave, we assume an enclosure similar to those at
the LET and OFFSPEC instruments on TS2 at ISIS, in the UK.  These are
hollow steel cans filled with borated paraffin wax.

The cost of concrete comes from ESS Conventional Facilities
Department, and is the price for reinforced concrete, cut and shaped
and installed.

The cost of raw metals come from the London Metal Exchange.

In all beamlines, it is anticipated to use at least three laminate
collimation blocks within the shielding bunker and/or curved sections,
to reduce the streaming of fast neutrons into the guide system
downstream.  These may be conceptually similar to the collimator on
the CHIPIR beamline at ISIS, or indeed much smaller, and spread
between multiple units.  Other names for these devices could include
``horse-collars'' or ``fast neutron scrapers''.  For the purposes of
this study, we assume these to be three units of 1\,m$^3$ of copper
with a small channel cut through the centre.

A summary of the items can be constructed as follows:

\begin{description}
\item[ISIS-TS2 Wax Cave] based on LET and OFFSPEC, for thermal beams
  out of line of sight.  Cost: 950 kGBP several years ago, rounded to
  1.5 M\euro.
\item[PSI Concrete Cave] based on BOA, for beams with direct line of
  sight: 2 M\euro.
\item[Heavy beam shielding] for beam areas within line of sight:
  approx. 7770 \euro $/m$.
\item[Light beam shielding] for thermal beams out of line of sight:
  approx. 2600 \euro $/m$.
\item[Laminate collimation blocks] initially costed as 100\% copper:
  47 k\euro per unit, 3 units per beam.
\item[T0/PPS Chopper]: only necessary for straight beamlines with line of
  sight of source, 750 k\euro.
\item[Heavy shutter]: only necessary for straight beamlines with line
  of sight of source, 750 k\euro.
\end{description}

Not included in this budget estimate are the ESS guide bunker and
target shielding, which absorb a large fraction of the radiation dose.
We only include shielding items outside the common shielding areas.
This has a potential cost-saving from the perspective of the
instrument project, since --- if the geometry loses line of sight
within the bunker --- the instrument only needs thermal beam shielding
in its budget.

The optical system costs are evaluated for the full length of the
beamline.  However, curving out of line of sight quickly, with a
radius of 1.5 km for example, increases the optical cost compared to a
gentle curvature of several km, because the neutron supermirrors need
to be engineered to reflect at larger grazing angles.  At still
tighter radii, of a few hundred metres, the neutron guides are
normally divided into several thin channels known as ``multi-channel
benders''.  These require more precision manufacturing, with a greater
surface that needs to be coated with supermirrors.  It should be clear
that, whilst curving out of line of sight quickly can reduce the
shielding cost, the optical cost is increasing against the shielding
cost saving.  These two costs are traded against each other in our
optimisation.

Optical component cost equations were provided to ESS as part of a
market survey, and these are commercial-in-confidence details that
cannot be widely shared.  However, they are comparable to costs of
items at similar facilities, as nothing in the market has changed
significantly in recent years.

\section{Requirements and Matched Optical Solutions}

\begin{figure}
  \includegraphics[width=\linewidth]{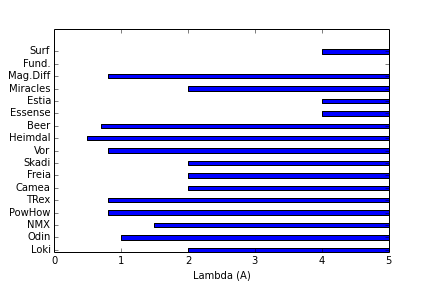}
  \caption{Wavelengths requested by each instrument on the reference suite at ESS.}
  \label{fig:LambdaSuite}
  \end{figure}

Instrument requirements were extracted initially from the ESS
Technical Design Report (TDR) \cite{ESS-TDR-2013} and gradually
refined with requirements described in the respective instrument
proposals. The minimum wavelength band of the instrument suite is
shown in figure \ref{fig:LambdaSuite}.  Here we can see that there are
three categories of instrument: those who are interested in
wavelengths at and just below 1 \AA{} and above; those interested in 2
\AA{} neutrons and above, and those interested only in cold neutrons
of 4 \AA{} wavelength and longer.

Subdividing into long (150 m) and medium (60 m) instruments is
informative.  The minimum wavelength band of the long instruments is
shown in figure \ref{fig:longBenderTransmission}, where we see that the long
instruments are split almost 50:50 into 2\AA{} and $\sim$1\AA{}
instruments.

For the medium-length instruments, the wavelength bands are shown in
figure \ref{fig:LambdaMediumSuite}.  In that subset, there are two
instruments requiring wavelengths around 1\,\AA{}, and a third
requiring 5\,\AA{}.

The shorter instruments follow a similar process, except with the
15\,m instruments who are considered to be so short that they are
either straight or use a multi-channel bender.

We can now begin to construct a critical performance/cost ratio
transport system as a standard, to compare with the baseline systems
from the instrument teams.  There are four curved guide systems to
consider: two sets of curved guides for the 150 metre instruments for
the two wavelength ranges; two sets of curved guides for the 60 metre
instruments for the two wavelength ranges.  Similarly, there are four
types of benders for the two lengths and two wavelength ranges.  The
150 metre instruments can have a 20 metre long bender within the
bunker, and the 60 metre instruments can have a 15 metre long bender,
and each of these is designed for different wavelength bands.

\subsection{Standard Guide Concept}
To compare with the proposed optical systems, a more minimalist
neutron guide concept was developed at ESS for these kinds of studies.
In this ``ESS Standard Guide'', elliptic geometries are replaced with
constant cross section geometries or ballistic geometries.
Supermirror $m$ values are capped at $m=4$ for 12 m long ballistic
focussing sections, and $m=1.5$ everywhere else.  Note that this
concept still provides more than 50\% of the neutron flux at 1\,\AA{} if
the guides are curved, due to the large radii involved with such long
guides.  The maximum dimensions of the guide are limited to below 4 cm
in the curved sections.

\subsection{Curved Guides}
A number of curved guide options will be considered.  The performance
estimates follow the work of Mildner
\cite{MILDNER-CURVED-GUIDE-ACCEPTANCE-DIAGRAMS} which is reliable for
the low-divergence parts of the guide systems with low-$m$ values,
which is the strategy in this case.

All the guides in the low cost options will be considered to have a
maximum size of 200 mm in horizontal and vertical extent.  The curved
guides have an additional constraint that the curved parts have a
width of 4 cm.

\begin{figure}
  \subfloat[]{
    \includegraphics[height=0.2\textheight]{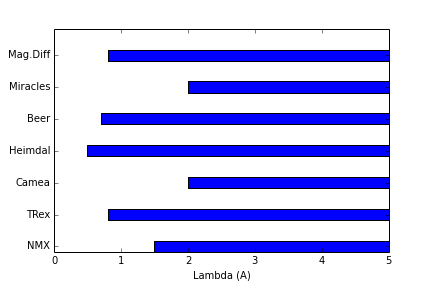}
  } \\
  \subfloat[]{
    \includegraphics[height=0.2\textheight]{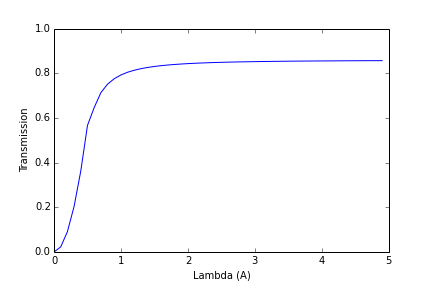}
  } \\
  \subfloat[]{
    \includegraphics[height=0.2\textheight]{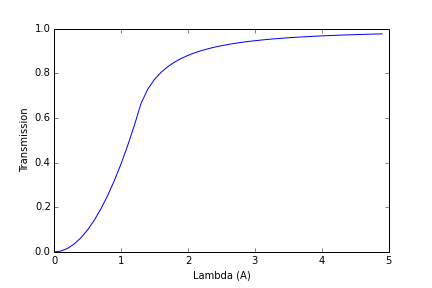}
  }
  \caption{\\(a) Wavelengths requested by the 150\,m long instruments
    at ESS. \\ (b) Transmission of a multi-channel bender designed to
    transmit 1\,\AA{} neutrons to the 150\,m long instruments at ESS.
    The specifcations of this bender are given in table
    \ref{tab:150mExpensiveBender} \\ (c) Transmission of a
    multi-channel bender designed to transmit 2\,\AA{} neutrons to the
    150\,m long instruments at ESS.  The specifcations of this bender
    are given in table \ref{tab:150mCheapBender}}.
      \label{fig:longBenderTransmission}
\end{figure}

\begin{figure}
    \subfloat[]{
      \includegraphics[height=0.2\textheight]{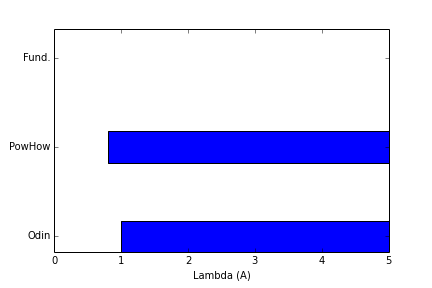}
    } \\
    \subfloat[]{
    \includegraphics[height=0.2\textheight]{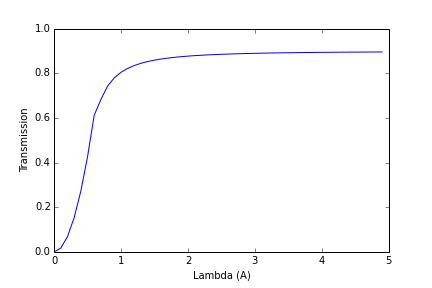}
    } \\
    \subfloat[]{
      \includegraphics[height=0.2\textheight]{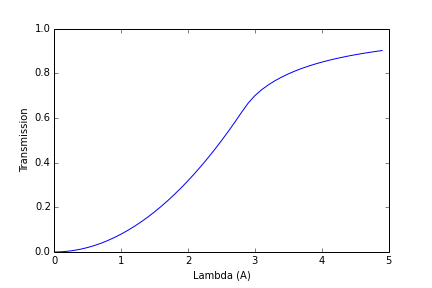}
    }
    \caption{\\ (a) Wavelengths requested by the 60\,m medium length
      instruments at ESS.  ``Fund'' corresponding to ``Fundamental
      physics'' requests 5\,\AA. \\ (b) (b) Transmission of a
      multi-channel bender designed to transmit 1\,\AA{} neutrons to the
      60\,m long instruments at ESS.  The specifcations of this bender
      are given in table \ref{tab:60mExpensiveBender}. \\ (c)
      Transmission of a multi-channel bender designed to transmit
      4\,\AA{} neutrons to the 60\,m long instruments at ESS.  The
      specifcations of this bender are given in table
      \ref{tab:60mCheapBender}.}
  \label{fig:LambdaMediumSuite}
\end{figure}

The parallel section of the guides is capped at $m$=1.5.  If we
restrict ourselves to guarantee the phase space homogeneity at
1.5\,\AA{} then, with 12\,m long compression/expansion sections at
the ends of the ballistic guides, we only require $m$=4 coatings.
This still provides a maximum beam divergence of 1$^\circ$.  The
transmission of this guide system is shown in figure
\ref{fig:longBenderTransmission}.  For the 60 metre guides, a higher
$m$ is required to get out of line of sight in a shorter distance.
The transmission of this guide system is shown in figure
\ref{fig:LambdaMediumSuite}.

\subsection{Multi-Channel Bender Options}
These items are necessary to get out of line of sight quickly in the
bunker.  The first solution matches the requirements of transmitting
1\,\AA{} neutrons, and is described in table
\ref{tab:150mExpensiveBender}.

The second version of this item is designed to transmit 2\,\AA{}
neutrons, and is described in table \ref{tab:150mCheapBender}.

\begin{table}
\caption{150 metre bender specifications for efficient transmission of
  1\,\AA{} neutrons.\label{tab:150mExpensiveBender}}
\begin{ruledtabular}
  \begin{tabular}{ll}
    {\bf Parameter}		& {\bf Value} \\
    \hline
    150 m bender width	& 4.0 cm \\ %
    150 m bender length     & 20 m   \\
    150 m channel width	& 0.5 cm\\
    150 m bender m		& 3.0\\
    150 m nchannels		& 8.0\\
    150 m bender radius	& 1250.0 m\\
    150 m transmission at 1\,\AA{}	&  80\% \\
    150 m Cost & 1.444 M\euro \\
  \end{tabular}
\end{ruledtabular}
\end{table}

\begin{table}
  \caption{150 metre bender specifications for efficient transmission
    of 2\,\AA{} neutrons.\label{tab:150mCheapBender}}
  \begin{ruledtabular}
      \begin{tabular}{ll}
        
            {\bf Parameter}		& {\bf Value} \\
            \hline
        150 m bender width		& 4.0 cm \\ %
        150 m bender length     & 20 m   \\
        150 m channel width		& 2 cm\\
        150 m bender m		& 2.5\\
        150 m nchannels		& 2\\
        150 m bender radius		& 1250.0 m\\
        150 m transmission at 2\AA{}		&  88\% \\
        150 m Cost & 341 k\euro\\
      \end{tabular}
      \end{ruledtabular}
\end{table}

\begin{table}
  \caption{60\,m bender for 1\,\AA{} neutrons.\label{tab:60mExpensiveBender}}
    \begin{ruledtabular}
\begin{tabular}{ll}  

    {\bf Parameter}		& {\bf Value} \\
\hline 
60 m bender width		& 3.0 cm \\ %
60 m bender length     & 15 m   \\
60 m channel width		& 0.5 cm\\
60 m bender m		& 3.0\\
60 m nchannels		& 6.0\\
60 m bender radius		& 937.5 m\\
60 m transmission at 1\AA{}		&  73\% \\
60 m Cost & 742 k\euro\\
\end{tabular}
\end{ruledtabular}
\end{table}

\begin{table}
  \caption{60\,m bender for 4\,\AA{} neutrons.\label{tab:60mCheapBender}}
    \begin{ruledtabular}
      \begin{tabular}{ll}
    {\bf Parameter}		& {\bf Value} \\
 \hline
60 m bender width		& 4.0 cm \\ %
60 m bender length     & 15 m   \\
60 m channel width		& 2 cm\\
60 m bender m		& 1.5\\
60 m nchannels		& 2\\
60 m bender radius		& 703.0 m\\
60 m transmission at 4\AA{}		&  85\% \\
60 m Cost & 193 k\euro \\
      \end{tabular}
      \end{ruledtabular}
\end{table}

\section{Suite Cost Totals\label{sec:SuiteCostTotals}}

\begin{figure}[htpb]
\begin{center}
\includegraphics*[width=\linewidth]{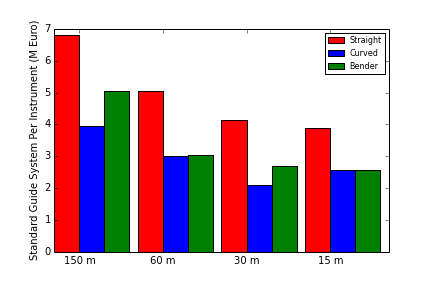}
\end{center}
\caption{Cost of each instrument using the standard ESS guide concept,
  to isolate the cost delta for the different curving options
  (therefore omitting the baseline suite).}
\label{fig:optionCosts}
\end{figure}

The cost per instrument as a function of instrument length is shown
for the standard instrument options in figure \ref{fig:optionCosts} on
page \pageref{fig:optionCosts}.
There it is clear that a general strategy of curving all the guides is
cheaper overall.  The cheapest possible method is using simple curved
guides, since the additional cost of multi-channel benders is not
offset by the cost savings in shielding outside the bunker.  On the
other hand, these marginal differences between the curved guides and
benders are a small increment to pay for likely improvements in the
instrument backgrounds.

Furthermore, we also see that the total cost delta for a straight
guide \emph{vs} a curved guide is 1-2 M\euro{} per instrument, on
average.  This seems approximately correct, considering the main cost
drivers in the cost delta, namely a $T0$ chopper, heavy shutter, and
enhanced instrument cave.

\begin{figure}[htpb]
\begin{center}
\includegraphics*[width=\linewidth]{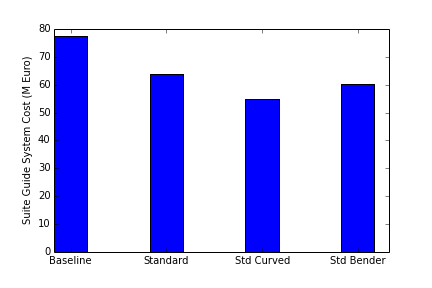}
\end{center}
\caption{Total cost for the instrument suite considered.}
\label{fig:suiteCost}
\end{figure}

\begin{figure}[htpb]
\begin{center}
\includegraphics*[width=\linewidth]{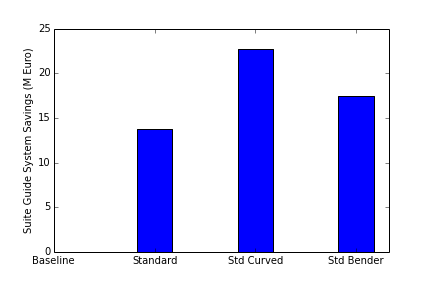}
\end{center}
\caption{Total cost saving for the instrument suite considered, for
  each of the options, relative to the baseline of as-proposed optics
  geometry.}
\label{fig:suiteCostSavings}
\end{figure}

The total instrument suite cost for the full suite is shown in figure
\ref{fig:suiteCost} on page \pageref{fig:suiteCost} for each option,
and the potential savings in figure \ref{fig:suiteCostSavings} on page
\pageref{fig:suiteCostSavings}.
In these figures, the following options are costed:
\begin{itemize}
\item ``Baseline'' which is the total of the instrument proposal
  costs, corrected where necessary (explanation in next paragraph)
\item ``Standard'' which is a reduction of the optical specifications
  towards a standard ESS guide
\item ``Std Curved'' which forces all beamlines to be curved out of
  line of sight, losing line of sight 50\% of the way down the
  beamline, in addition to following the ESS standard guide design
\item ``Std Bender'' which follows the ESS standard guide design and
  has a multi-channel bender to lose line of sight in the bunker.
\end{itemize}
One can see that, relative to the baseline, the standard curved guide
concept is likely to save 24 M\euro{} across the suite, with minimal
impact on the instrument performance.

\section{Costing Validation}
%
%

For any publicly funded project it is important to get an as good as
possible cost indication early and refine estimations as soon as more
detailed information becomes available.  The instruments described in
\cite{ESS-TDR-2013} allowed for an indicative cost, which is part of
the total ESS construction cost of 1.843\,B{\euro} established in
2013.

Over the last 3 years, the instrument concepts were refined primarily
in terms of scientific requirements but not so much in terms of costs.
On the other hand it is the central facility’s responsibility to
manage the available budgets as efficiently as possible to maximise
the scope for and involvement of in-kind partners throughout Europe.
In order refine the cost of neutron instrument components (aka
beamline components or instrument components or simply components), a
workshop was held in early 2014, which involved partners from ISIS in
the UK, PSI from Switzerland, and JCNS from Germany, all of which have
an extensive and excellent track record of building world-class
neutron instruments .

During the workshop, a list of around 20 components with indicative
costs --- based on recent projects such as ISIS TS-II \cite{ISIS-TS2} or
the JCNS spin-echo instrument BL-15 at SNS \cite{JCNS-SNS} --- was
established, which included e.g. shutters, benders, instrument caves
but also services for installation or staff/labour costs
\cite{ESS-CostCatWorkshop}.  Although the scope of the instrument
programme was not defined at that time, the components were used to
\begin{enumerate}
\item Better estimate the costs of the TDR reference suite,
\item To gradually refine costs as instrument concepts became better
  defined, and
\item To independently assess costs to obtain the best
  ``value-for-money'' for all the partner countries involved in the
  instrument build programme, and consequently the tax payers within
  the partner countries.
\end{enumerate}
  
\begin{figure}[htpb]
\begin{center}
\includegraphics*[width=\linewidth]{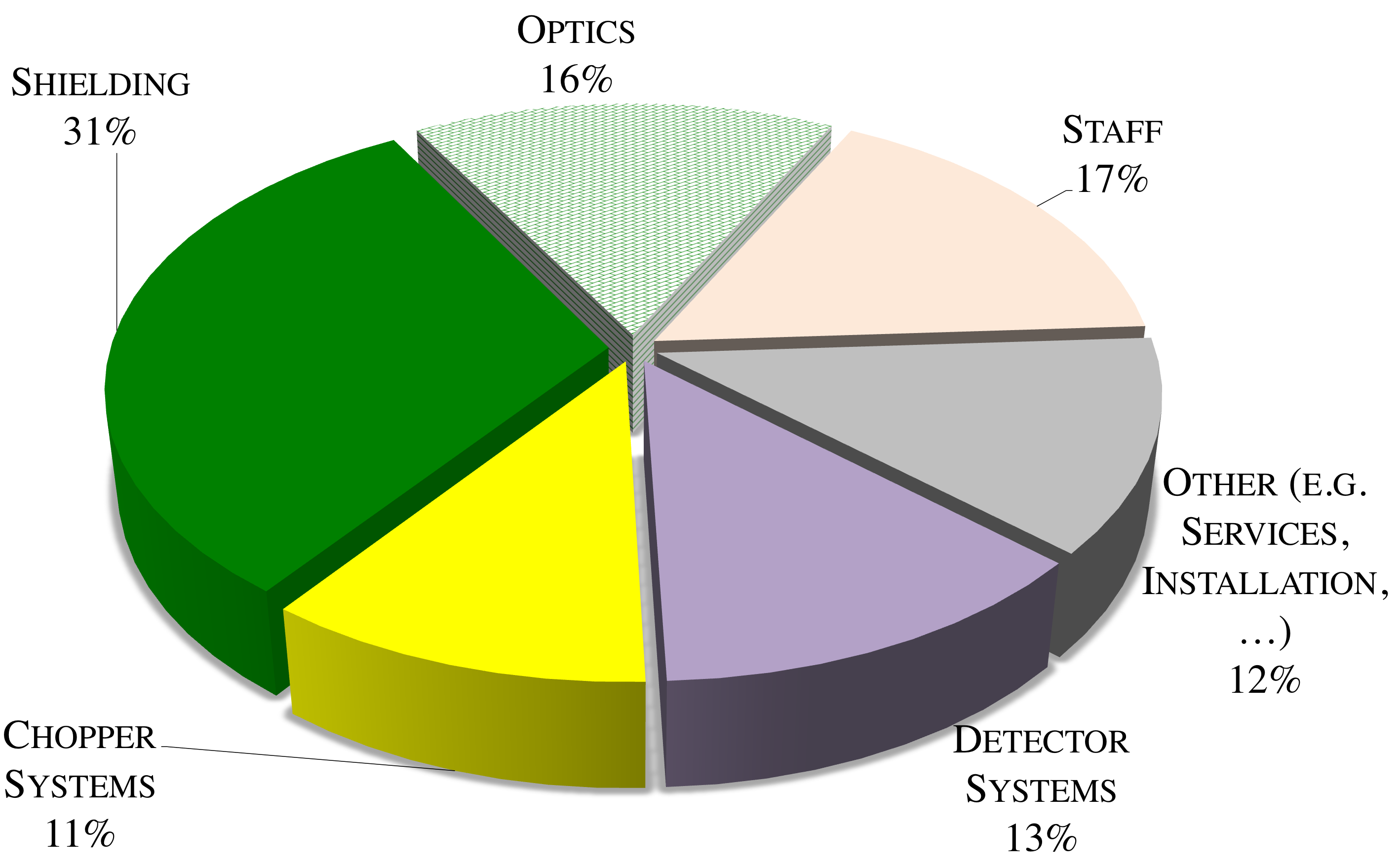}
\end{center}
\caption{Approximate cost distribution of components for neutron
  scattering instruments considered to be built at the ESS.}
\label{fig:InstrumentCostDistribution}
\end{figure}

From \cite{ESS-CostCatWorkshop}, a cost distribution for instruments
as shown in fig. \ref{fig:InstrumentCostDistribution} can be obtained.
While --- for good reason --- a lot of focus is currently on
e.g. detector systems for ESS instruments due to the difficulty to
acquire large quantities of $^3$He gas \cite{ESS-Detector-Strategy},
it may come as a surprise that around 47\% of the total costs fall
into the categories ``Shielding'' and ``Optics'', with 31\% and 16\%,
respectively. For the instruments described in the TDR --- with adjustments to
consider improvements in the scientific requirements/specifications
--- one would arrive at an overall cost of around 290\,M{\euro}.
While the reference suite comprises 22 instruments, the current scope
of the instrument build programme is 16 (it is worthwhile noting that
the 16 instruments are part of the ESS Construction Programme; plans
have been developed and will be refined over the next 2-3 years to
arrive at 22 public instruments described in \cite{ESS-TDR-2013}).

For a suite of 16 instruments one would estimate costs of
211\,M{\euro}. As indicated by
fig. \ref{fig:InstrumentCostDistribution} the majority of the costs
are associated with shielding/optical components, and amongst those
components the main cost drivers are instrument caves (initially
valued at 2.5\,M{\euro}) and neutron guides (initially valued at
~30k{\euro}/m; including not only the glass components but also
required shielding).  A simplistic approach ignoring any cost-benefit
analysis would require for only instrument caves and guide systems of
16 instruments a budget of around 83\,M{\euro}, which is consistent
within an error of around 10\% with the baseline cost indicated in
fig. \ref{fig:suiteCostSavings}.  The arguments provided earlier in
terms of optimising e.g. instrument caves as a function of
`curved/straight' immediately changes cost requirements for only
instrument caves from around 40\,M{\euro} to 26\,M{\euro} thus
giving a potential cost saving of 14\,M{\euro}.  Similar arguments
can be applied for different guide geometries, curvatures resulting in
approx. 1\,M\euro{} per instrument, and a total of 16\,M{\euro} $+$
14\,M{\euro}$=$ 30\,M{\euro}, which is --- again within a 10\%
error --- consistent with results discussed in the context of
fig. \ref{fig:suiteCostSavings}.

In terms of project management one could also use the results
indicated in fig. \ref{fig:suiteCostSavings} to perform an indicative
PERT (e.g. \cite{PERT-ESTIMATION}) estimation with
\begin{itemize}
\item the {\it{Std Curved option}} saving being the optimistic estimation
\item the {\it{Standard option}} being the pessimistic estimation
\item the {\it{Std Bender option}} being the best guess estimation
\end{itemize}

Usually, a PERT analysis is used to regulate optimism and pessimism in
schedule estimates, but we can also apply the same logic here for
cost option extrema, resulting in a potential saving of 17\, M\euro.

\subsection{Approximate Benchmarking}
For the moment it is assumed that the cost per instrument can be
reduced from 14.5\,M\euro\, to 13.5\,M\euro\, using the more
conservative PERT estimation. At ESS the average length of an
instrument is 90\,m due to the long pulse design. At short pulse
spallation sources such as SNS or ISIS instruments are on average
around 50\,m long, resulting in an average cost of around
12.2\,M\euro$_{2015}$.

ISIS' recent TS2 project was established with an budget of
145\,M\pounds, out of which 100.5\,M\pounds\, were allocated to the
core project and 27\,M\pounds\, for the instrument project with the
remaining 17.5\,M\pounds\, for instruments coming from the EU and
collaborating European countries \cite{ISIS-TS2-BC}, a similar model
that is used for the phase 2 upgrade of TS2 \cite{ISIS-TS2}.  The
initial scope of TS2 included 7 instruments or an average cost of
6.4\,M\pounds. A conservative estimation of inflation would assume a
cost increase of around 20\%\, since 2005 \cite{EuropeanHCIP}
resulting in 7.7\,M\pounds$_{2015}$. Over the last years the average
exchange ration between \euro\, and \pounds \, was 1:1.28
\cite{HistoricConverter} giving 9.8\,M\euro$_{2015}$\, for an average
instrument of around 50 m length. The main difference between the
12.2\,M\euro$_{2015}$\, and the 9.8\,M\euro$_{2015}$\, can easily be
explained by a slightly higher-level of complexity of ESS instruments
due to the different source characteristics, but also due to the fact
that ESS is a green field site. Also, it can be assumed that
additional effort, in particular with regard to installation, is
required to due to the in-kind nature of the project.


While all the numbers are only indicative --- details matter after all ---
they nevertheless show that proper strategies and analyses such as
benchmarking, component breakdown, cost-benefit analysis or efficient
standardised technical solutions via optimised non-recurrent
engineering designs have the potential to optimise costs, which is not
a surprise in itself.  Although the focus here is on optical and
shielding components any cost optimisation has two major consequences:
obviously, it reduces overall costs, which is in the interest of the
project governance and the funding agencies; but secondly, because
standardised systems are proposed with almost no or minimal impact on
performance, costs related to installation, maintenance and spare
parts are being reduced, which in turn will lower the operational
costs.  As the operational costs are estimated to be typically 4 times
the initial investment cost in case of a scientific facility, this has
a significant effect over the lifetime of the facility.

It may actually be more appropriate to refer to a cost optimisation of
{\it{x\,M\euro}} \,because the aim is not to minimise the costs by all
means but to rather use the available budget more intelligently and
focused by e.g. redistributing costs into areas such as computing,
installation or contingency if required, but also to compensate for
any shortfalls of the initial assumptions when creating the list of
components. It is also worthwhile noting that at this point no
significant de-scoping of instruments in terms of their scientific
performance or functionality is discussed.  This is an alternative
project management tool that, whilst popular and utilised in large
scale scientific projects, nonetheless can have a devastating impact
on the scientific output of a facility, and the methodology presented
here should probably be used \emph{before} removing scope from the
instrument projects themselves.

%
%

\section{Conclusions}
We have used neutron optics geometry combined with a cost-benefit
analysis balancing shielding costs, to propose an optimisation of the
ESS instrument suite budget.  Cost savings in the order of 20\,M\euro
\, are identified compared to the baseline budget as defined in the
adjusted instrument proposal costs.  This cost saving has been
compared with instrument costs at existing facilities, and found to
be within reasonable agreement.  While this approach is common in
industry, it is perhaps less widely used in scientific projects, and
the management challenge to unlock these savings, at any facility, is
to establish a strong central coordination mechanism: to supply
standard concepts and systems such as the ones described here; to
provide a core team to manage the suite optimisation; and to intervene
when deviations from the standards are being pursued, particularly in
the cases where small performance increases are accompanied by large
cost impacts.

\begin{acknowledgments}
The information and support given by our partners and colleagues from
ISIS, PSI and JCNS, in particular with regard to establishing a cost
baseline for instrument components, is greatly appreciated.  We are
also extremely grateful for the feedback given at various ESS project
reviews on this work, particularly K. Herwig, D. Mildner, E. Iverson,
M. Fitzsimmons, S. Kennedy, T. Sato and T. Krist.  Useful discussions
with the rest of the Neutron Optics and Shielding Group during
2013-2014 were invaluable, particularly C. Zendler, D. Martin
Rodriguez, D. DiJulio, N. Cherkashyna, and C. Cooper Jensen.

\end{acknowledgments}


\end{document}